\documentclass[12pt,letterpaper]{article}
\usepackage{jheppub}
\usepackage{verbatim}
\usepackage{graphicx,times}
\usepackage{amsmath,amssymb,latexsym,float}
\usepackage{enumerate}

\numberwithin{equation}{section}
\newcommand{\norm}[1]{\left\Vert#1\right\Vert}

\newcommand{\R}{\text{\fontshape{n}\selectfont I\kern-.42exR}}

\newcommand{\1}{\text{\fontshape{n}\selectfont 1\kern-.56exl}}

\newcommand{\pslash}{\slash\!\!\!\!\!~p}

\title{Lattice QCD with QCDLAB}

\author{Artan Bori\c{c}i}
\affiliation{University of Tirana\\
             Blvd. King Zog I\\
             Tirana\\
             Albania
}

\emailAdd{borici@fshn.edu.al}

\abstract{QCDLAB is a set of programs,  written in GNU Octave, for lattice QCD computations. Version 2.0 includes the generation of configurations for the SU(3) theory, computation of rectangle Wilson loops as well as the low lying meson spectrum with Wilson fermions. Version 2.1 includes also the computation of the low lying meson spectrum using minimally doubled chiral fermions. In this paper, we give a brief tutorial on lattice QCD computations using QCDLAB.

\vfill
}

\begin{document}
\maketitle

\section{Introduction}

Quantum Chromodynamics (QCD) is the theory of strong interactions. QCD has an ultraviolet fixed point at vanishing coupling constant, a property which was first demonstrated in the perturbative formulation by Gross and Wilczek as well as by Politzer \cite{gross_wilczek,politzer}. A year later, Wilson was able to formulate QCD non-perturbatively \cite{wilson}. He showed, that in the strong coupling regime, QCD is confining, meaning that the potential between two static charges grows linearly with the separation of charges. Later that year, Kogut and Sussking extended the non-perturbative formulation in the Hamiltonian formalism \cite{wilson,kogut_susskind}. It was immediately clear that a direct evaluation of QCD path integral was only possible using Monte Carlo simulations. Creutz was the first to show numerically that the weak and strong regimes are in the same phase in four dimensions \cite{creutz}. Since then, lattice QCD has grown into a separate numerical discipline and has delivered results of growing accuracy. This development was possible from the exponential increase of computing power and more efficient algorithms.

In this paper we deal with the basic technology at the bottom of any contemporary lattice computation without going into the details that make lattice QCD confront experiment as well as predict physics beyond the Standard Model. Lattice QCD is a collaborative project, and as such, may not be brought into one review paper without missing a single contribution. Here we profit from the QCDLAB programs which is a small set of short programs that allows one to ilustrate the basic properties of QCD without getting bogged down into the details of advanced computing technology and associated software and algorithms. In contrast to other sotware, QCDLAB maps linear operators of QCD to linear operators of the GNU Octave language \cite{octave}. Although GNU Octave is an interpreted language, linear operators are precompiled. This property enables very efficient coding as well as minimal run times.

However, GNU Octave is a one-threaded software and runs in one computing core only. Therefore, QCDLAB usage is limited to moderate lattices. It is possible however to include multi-threaded C++ libraries such that the programs run in multiple cores. Writing dedicated libraries of this sort will drive the QCDLAB project out of the original aim of keeping the programing effort small. Nonetheless, Octave is a language in development and is likely to include in the future multi-threaded linear algebra libraries.

In summary, QCDLAB serves three purposes: teaching, learning as well as algorithm prototyping. The latter helps developing a complex software by testing the basic idea of a new algorithm on GNU Octave using QCDLAB codes. QCDLAB programs, version 2.0, as well as this document are available at:
\begin{center}
\verb+https://sites.google.com/site/artanborici/qcdlab+
\end{center}
It is licensed under the GNU General Public License v3. The present document serves as a user guide of QCDLAB as well as an illustration of basic calculations in lattice QCD.

\section{QCD data}

The space-time world in lattice QCD is taken to be a four dimesnional regular lattice. At each lattice site go out four directed links, as ilustrated below in the case of two dimensions. The lattice sites are numberd in a lexicographical order.

\begin{center}
\includegraphics[scale=0.5]{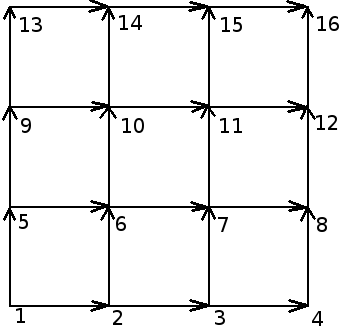}
\end{center}

\noindent
The basic degree of freedom on the lattice is the gauge field. Lattice gauge fields are SU(3) group elements in the fundamental representation, i.e. $3\times 3$ complex unitary matrices with determinant one. We associate one such element to each directed link on the lattice. If $i$ and $i+\hat\mu$ label two neighboring lattice sites along the direction $\mu\in\{1,2,3,4\}$ the associated link is denoted by $U_{\mu,i}$ as in the figure.

\begin{center}
\includegraphics[scale=0.3]{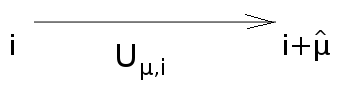}
\end{center}

\subsection{Lattice configurations}

The collection of all lattice links is a lattice configuration. The basic linear operator in lattice QCD is the matrix of such a configuration along the direction $\mu$:
$$
U_\mu=\begin{pmatrix}U_{\mu,1}& & &\\& U_{\mu,2} & &\\& &\ddots&\\&&&U_{\mu,N}\end{pmatrix}\ ,
$$
where $N$ is the total number of lattice sites and the ordering is lexicographical. If $N_1,N_2,N_3,N_4$ are the number of sites along each direction, the total number of sites is $N=N_1N_2N_3N_4$. Note that $U_\mu$ is a block diagonal matrix with blocks of $3\times 3$ size.

QCDLAB follows the same data structure. For example, a random gauge field, which is appropriate for QCD at strong copuling can be generated by the following routine:
\begin{small}
\begin{verbatim}
function U=RandomGaugeField(N);
%
U1=[]; U2=[]; U3=[]; U4=[];
for k=1:N;
  [u1,R]=qr(rand(3)+sqrt(-1)*rand(3)); u1(:,3)=u1(:,3)/det(u1);
  [u2,R]=qr(rand(3)+sqrt(-1)*rand(3)); u2(:,3)=u2(:,3)/det(u2);
  [u3,R]=qr(rand(3)+sqrt(-1)*rand(3)); u3(:,3)=u3(:,3)/det(u3);
  [u4,R]=qr(rand(3)+sqrt(-1)*rand(3)); u4(:,3)=u4(:,3)/det(u4);
  U1=[U1,u1];
  U2=[U2,u2];
  U3=[U3,u3];
  U4=[U4,u4];
end
% form sparse matrices
[I,J]=find(kron(speye(N),ones(3)));
u1=sparse(I,J,U1,3*N,3*N);
u2=sparse(I,J,U2,3*N,3*N);
u3=sparse(I,J,U3,3*N,3*N);
u4=sparse(I,J,U4,3*N,3*N);
U=[u1,u2,u3,u4];
\end{verbatim}
\end{small}
Note the sequential loop creating the gauge fields one by one. It is the only instance where QCDLAB uses such a loop in connection to degrees of freedom. This routine is called only once, usually at the begining of the simulation code. The user can avoid its call by simply starting with identity matrices and create random gauge fields using the simulation routine. This is controlled by the \verb+GaugeField+ function. Note also that the set of four gauge fields is stored in a single matrix \verb+U=[u1,u2,u3,u4]+.

\subsection{Shift operators}

An important operator on the lattice is the permutation operator that shifts lattice sites along the positive direction $\mu$:
$$
T_\mu=\begin{pmatrix}&1&&&&&\\&&&1&&&\\&&&&&\ddots&\\1&&&&&&\end{pmatrix}\ .
$$
In total we have four such operators, one for each direction. In four dimensions these are built using Kronecker products with identity matrices $I_{N_1},I_{N_2},I_{N_3},I_{N_4}$:
\begin{eqnarray*}
E_1&=&I_4\otimes I_3\otimes I_2\otimes T_1\otimes I_3\\
E_2&=&I_4\otimes I_3\otimes T_2\otimes I_1\otimes I_3\\
E_3&=&I_4\otimes T_3\otimes I_2\otimes I_1\otimes I_3\\
E_4&=&T_4\otimes T_3\otimes I_2\otimes I_1\otimes I_3\ .
\end{eqnarray*}
Note that the extra Kronecker product with the identity $3\times 3$ matrix is neccessary in order to accomodate the space of gauge fields. The following routine creates the required operators:
\begin{small}
\begin{verbatim}
function E=ShiftOperators(N1,N2,N3,N4);
% Shift operators
p1=[N1,1:N1-1]; p2=[N2,1:N2-1]; p3=[N3,1:N3-1]; p4=[N4,1:N4-1];
I1=speye(N1); I2=speye(N2); I3=speye(N3); I4=speye(N4);
T1=I1(:,p1); T2=I2(:,p2); T3=I3(:,p3); T4=I4(:,p4);
e1=kron(I4,kron(I3,kron(I2,kron(T1,speye(3)))));
e2=kron(I4,kron(I3,kron(T2,kron(I1,speye(3)))));
e3=kron(I4,kron(T3,kron(I2,kron(I1,speye(3)))));
e4=kron(T4,kron(I3,kron(I2,kron(I1,speye(3)))));
E=[e1,e2,e3,e4];
\end{verbatim}
\end{small}
Like in the case of gauge fields the set of four shift operators is stored in a single matrix \verb+E=[e1,e2,e3,e4]+.

\subsection{Wilson action}

With the above operators we can write down the action of the SU(3) lattice theory as proposed by Wilson:
$$
S_{\text{gauge}}(U)=-\frac{1}{g^2}\sum_{\mu\nu}\text{tr~}(U_\mu E_\mu)(U_{\nu} E_\nu)(U_\mu E_\mu)^*(U_\nu E_\nu)^*\ ,
$$
where the star symbolises the matrix Hermitian conjugation, the trace is taken in the $3N$ dimensional space and $g$ is the bare coupling constant of the theory. In lattice gauge theory it is standart to use the inverse coupling constant:
$$
\beta=\frac{6}{g^2}\ .
$$
An important observable in QCD is the Wilson loop. The elementary Wilson loop, called {\it plaquette}, is the colorless product of gauge fields around an elementary suqare:
$$
{\cal P}_{\mu\nu,i}=\frac{1}{3}\text{Re tr}_{C}U_{\mu,i}U_{\nu,i+\hat\mu}U_{\mu,i+\hat\nu}^*U_{\nu,i}^*\ ,
$$
where the trace in performed in the color space.

\begin{picture}(100,180)(0,0)
\put(150, 40){\vector( 1, 0){100}}
\put(250, 40){\vector( 0, 1){100}}
\put(250,140){\vector(-1, 0){100}}
\put(150,140){\vector( 0,-1){100}}
\put(150, 30){$i$}
\put(250, 30){$i+\hat\mu$}
\put(250,140){$i+\hat\mu+\hat\nu$}
\put(120,140){$i+\hat\nu$}
\put(190, 20){$U_{\mu,i}$}
\put(260, 80){$U_{\nu,i+\hat\mu}$}
\put(190,150){$U_{\mu,i+\hat\nu}^*$}
\put(120, 80){$U_{\nu,i}^*$}
\end{picture}

\noindent
It is straightforward to see that the product of matrices in the Wilson action can be written as a sum over all plaquettes: 
$$
S_{\text{gauge}}(U)=-\beta\sum_{i,\mu>\nu}P_{i,\mu\nu}\ .
$$
The sum in the right hand side can be computed in QCDLAB by the following routine:
\begin{small}
\begin{verbatim}
function p=Plaquette(U);
% computes unnormalised plaquette
p=0;
%globals
global beta N E
for mu=1:4;
for nu=mu+1:4;
  E1=E(:,(mu-1)*3*N+1:mu*3*N); E2=E(:,(nu-1)*3*N+1:nu*3*N);
  U1=U(:,(mu-1)*3*N+1:mu*3*N); U2=U(:,(nu-1)*3*N+1:nu*3*N);
  U1=U1*E1; U2=U2*E2;
  p=p+real(trace(U1*U2*U1'*U2'));
end
end
\end{verbatim}
\end{small}
It follows directly the definition of the action in terms of sparse matrices, which makes the computation very efficient. Since the color trace is unnormalized in the routine the Wilson action is computed by calling \verb+-beta/3*Plaquette(U)+.

\subsection{Dirac operator}

Let $D(U)$ be the Wilson formulation of the Dirac operator describing one quark flavor with bare mass $m$ in the background gauge field configuration $U$:
$$
D(U)=(m+d)I-\frac{1}{2}\sum_{\mu}[(U_\mu E_\mu)\otimes(1-\gamma_\mu)+(U_\mu E_\mu)^*\otimes(1+\gamma_\mu)]\ ,
$$
where $d=4$ and $\gamma_\mu$ are $4\times 4$ Dirac matrices obeying the Dirac-Clifford algebra in the Euclidean signature:
$$
\{\gamma_\mu,\gamma_\nu\}=\delta_{\mu\nu}\ .
$$
We assume also periodic boundary conditions in each direction. To see why it works we set $U_\mu=I$ and go to momentum space, in which case $E_\mu(p)=e^{ip_\mu}$ and therefore:
$$
D(p)=m+i\sum_\mu\gamma_\mu\sin p_\mu+\sum_\mu(1-\cos p_\mu)\ .
$$
It is clear that for small momenta $D(p)\rightarrow m+i\pslash+p^2/2+O(\pslash^3)$, whereas at other corners of the Brillouin zone there are 15 additional heavy flavors with masses $m+2,m+4,m+6,m+8$ and spin structure described by different sets of gamma-matrices. Therefore, at small momenta, Wilson fermions describe a single flavor of fermions and break chiral symmetry even at $m=0$ by the $p^2/2$ term. We will discuss this issue further in later sections. Here is the routine that implements Wilson fermions:
\begin{small}
\begin{verbatim}
function A=Wilson(U,mass);
% Constructs Wilson-Dirac lattice operator
%global mass N N1 N2 N3 N4 E1 E2 E3 E4 GAMMA5
N1=6; N2=6; N3=6; N4=12; N=N1*N2*N3*N4;
% gamma matrices
gamma1=[0, 0, 0,-i;  0, 0,-i, 0;  0, i, 0, 0;  i, 0, 0, 0];
gamma2=[0, 0, 0,-1;  0, 0, 1, 0;  0, 1, 0, 0; -1, 0, 0, 0];
gamma3=[0, 0,-i, 0;  0, 0, 0, i;  i, 0, 0, 0;  0,-i, 0, 0];
gamma4=[0, 0,-1, 0;  0, 0, 0,-1; -1, 0, 0, 0;  0,-1, 0, 0];
% Projection operators
P1_plus=eye(4)+gamma1; P1_minus=eye(4)-gamma1;
P2_plus=eye(4)+gamma2; P2_minus=eye(4)-gamma2;
P3_plus=eye(4)+gamma3; P3_minus=eye(4)-gamma3;
P4_plus=eye(4)+gamma4; P4_minus=eye(4)-gamma4;
% Shift operators
p1=[N1,1:N1-1]; p2=[N2,1:N2-1]; p3=[N3,1:N3-1]; p4=[N4,1:N4-1];
I1=speye(N1); I2=speye(N2); I3=speye(N3); I4=speye(N4);
T1=I1(:,p1); T2=I2(:,p2); T3=I3(:,p3); T4=I4(:,p4);
E1=kron(kron(kron(kron(T1,I2),I3),I4),speye(3));
E2=kron(kron(kron(kron(I1,T2),I3),I4),speye(3));
E3=kron(kron(kron(kron(I1,I2),T3),I4),speye(3));
E4=kron(kron(kron(kron(I1,I2),I3),T4),speye(3));
%
U1=U(:,0*3*N+1:1*3*N);
U2=U(:,1*3*N+1:2*3*N);
U3=U(:,2*3*N+1:3*3*N);
U4=U(:,3*3*N+1:4*3*N);
%
% Upper triangular
A=  kron(U1*E1,P1_minus);
A=A+kron(U2*E2,P2_minus);
A=A+kron(U3*E3,P3_minus);
A=A+kron(U4*E4,P4_minus);
% Lower triangular
A=A+kron(U1*E1,P1_plus)';
A=A+kron(U2*E2,P2_plus)';
A=A+kron(U3*E3,P3_plus)';
A=A+kron(U4*E4,P4_plus)';
A=(mass+4)*speye(12*N)-0.5*A;
\end{verbatim}
\end{small}

\subsection{A first algorithm}

One special task in QCDLAB is the exponentiation of $su(3)$ algebras. The concrete form of an $su(3)$ algebra associated to a SU(3) gauge field in the fundamental representation is a $3\times 3$ anti-Hermitian traceless matrix. We have the following task: given a block diagonal matrix $P_\mu$ of order $3N$ with non zero $su(3)$ algebra blocks we would like to compute the gauge field matrix:
$$
U_\mu=e^{P_\mu}
$$
without using loops over the lattice sites. Here is an algorithm that completes this task:
\begin{small}
\begin{verbatim}
function U=Exp_su3(P);
% exponentiate su(3) algebras
% using power expansion and Horner's algorithm
global N
%
P1=P(:,0*3*N+1:1*3*N);
P2=P(:,1*3*N+1:2*3*N);
P3=P(:,2*3*N+1:3*3*N);
P4=P(:,3*3*N+1:4*3*N);
%
Id=speye(max(size(P1)));
u1=Id; u2=Id; u3=Id; u4=Id;
n=24;
for k=n:-1:1;
 u1=Id+P1*u1/k;
 u2=Id+P2*u2/k;
 u3=Id+P3*u3/k;
 u4=Id+P4*u4/k;
end
U=[u1,u2,u3,u4];
\end{verbatim}
\end{small}
It is an implemetation of the exponential power expansion:
$$
e^{P_\mu}=\sum_{k=1}^n\frac{P_\mu^k}{k!}+O\left[P_\mu^{(n+1)}\right]
$$
truncated at order $n=24$ using the Horner algorithm. The order is chosen such that the resulting gauge fields are SU(3) matrices in the working precision of GNU Octave. If in doubt, the user should use the routine \verb+Unitarity_check+. There are more efficient implementations if we were to write the routine in C++. In this case one can exponentiate $su(3)$ algebras one at a time using the algorithm behind the \verb+expm+ function of the GNU Octave.

\section{QCD path integral}

In this paper we focus in the simulation of pure Yang-Mills theory. Simulation of lattice QCD in this approximation, known as the quenched approximation, neglects screening coming from quark-antiquark pairs. It delievers very fast QCD properties such as the linear rising potential and hadron spectrum. Therefore, our task is the evaluation of the path integral:
$$
Z=\int\prod_{\mu,i}dU_{\mu,i}~e^{-S_{\text{gauge}}(U)}\ ,
$$
where $dU_{\mu,i}$ denotes the the SU(3) group integration measure, which is asummed to be gauge invariant. Its concrete form is unimportant in the algorithms used in QCDLAB.

\subsection{Hybrid Monte Carlo Algorithm}

The HMC algorithm \cite{hmc} starts by introducing $su(3)$ conjugate momenta matrices $P_\mu$ to gauge fields. Gauge field configuations are generated by integrating classical field equations with Hamiltonian:
$$
{\cal H}(P,U)=-\frac{1}{4}\text{tr~}\sum_{\mu}P_\mu^2-\frac{\beta}{6}\sum_{\mu\nu}\text{tr~}(U_\mu E_\mu)(U_{\nu} E_\nu)(U_\mu E_\mu)^*(U_\nu E_\nu)^*\ .
$$
The extra one half in the normalization of the kinetic energy comes from the normalization of Gell-Mann matrices, which are adopted as $su(3)$ algebra generators in the calculation of momentum matrices. The kinetic energy is computed by the following routine:
\begin{small}
\begin{verbatim}
function y=T(P);
% computes the kinetic energy of H
global N
p1=P(:,0*3*N+1:1*3*N);
p2=P(:,1*3*N+1:2*3*N);
p3=P(:,2*3*N+1:3*3*N);
p4=P(:,3*3*N+1:4*3*N);
y=-(trace(p1^2)+trace(p2^2)+trace(p3^2)+trace(p4^2))/4;
y=real(y);
\end{verbatim}
\end{small}
The first equation of motion is:
$$
\dot U_{\mu}=P_{\mu}U_{\mu}\ .
$$
Since ${\cal H}$ is an integral of motion, the second equation is derived by the equation:
$$
0=\dot{\cal H}=-\frac{1}{4}\sum_{\mu} \text{tr~}P_{\mu}\dot P_{\mu}
-\frac{\beta}{6}\sum_{\mu\nu}(\dot U_{\mu} E_\mu)(U_{\nu} E_\nu)(U_\mu E_\mu)^*(U_\nu E_\nu)^*+\text{h.c.}~~\ .
$$
Substituting for $\dot U_{\mu}$ the first equation of motion:
$$
0=\dot{\cal H}=-\frac{1}{2}\sum_{\mu} \text{tr~}P_{\mu}\left\{\frac{1}{2}\dot P_{\mu}+\frac{\beta}{3}\sum_{\nu(\neq\mu)}\left[{\cal P}_{\mu\nu}^{(1)}+{\cal P}_{\mu\nu}^{(2)}\right]\right\}+\text{h.c.}
$$
with ${\cal P}_{\mu\nu}^{(1)}=(U_{\mu} E_\mu)(U_{\nu} E_\nu)(U_\mu E_\mu)^*(U_\nu E_\nu)^*\ ,~{\cal P}_{\mu\nu}^{(2)}=(U_{\mu} E_\mu)(U_\nu E_\nu)^*(U_\mu E_\mu)^*(U_{\nu} E_\nu)$ one gets the second equation of motion:
$$
\frac{1}{2}\dot P_{\mu}=-\frac{\beta}{3}\sum_{\nu(\neq\mu)}\left[{\cal P}_{\mu\nu}^{(1)}+{\cal P}_{\mu\nu}^{(2)}\right]\ .
$$
Since ${\cal P}_{\mu\nu}$ matrices are $1\times 1$ loops around adjecent plaquettes that share a common link, they are block diagonal matrices of $3\times 3$ blocks. However, these blocks are not guaranted to be $su(3)$ valued. Therefore, the force exerted at the gauge field $U_{\mu,i}$ is the traceless anti-Hermitian part of $\dot P_{\mu,i}$:
$$
F_{\mu,i}=\frac{1}{2}(\dot P_{\mu,i}-\dot P_{\mu,i}^*)-\frac{1}{3}\text{tr}_C~\frac{1}{2}(\dot P_{\mu,i}-\dot P_{\mu,i}^*)\ .
$$
The force is implemented in the following routine:
\begin{small}
\begin{verbatim}
function F=Force_su3(U);
%globals
global beta N E
F=[];
for mu=1:4;
  M=sparse(zeros(3*N));
  for nu=1:4;
    if (mu~=nu),
      E1=E(:,(mu-1)*3*N+1:mu*3*N); E2=E(:,(nu-1)*3*N+1:nu*3*N);
      U1=U(:,(mu-1)*3*N+1:mu*3*N); U2=U(:,(nu-1)*3*N+1:nu*3*N);
      U1=U1*E1; U2=U2*E2;
      M=M+U1*U2*U1'*U2'+U1*U2'*U1'*U2;
    endif
  end
  f=M-M';
% subtract trace
  diag_f=diag(f); tr_f=sum(reshape(diag_f,3,N));
  tr_f=kron(transpose(tr_f),ones(3,1));
  f=f-sparse(diag(tr_f))/3;
  F=[F,f];
end
F=-beta/3*F;
\end{verbatim}
\end{small}

Having the equations of motion the next step is to build a trajectory using the leapfrog integration scheme:
\begin{eqnarray*}
U_{\mu}(t+\frac{\Delta t}{2})&=&e^{P_{\mu}(t)\Delta t/2}U_{\mu}(t)\\
P_{\mu}(t+\Delta t)&=&P_{\mu}(t)+F_{\mu}(t+\Delta t/2)\Delta t\\
U_{\mu}(t+\Delta t)&=&e^{P_{\mu}(t+\Delta t)\Delta t/2}U_{\mu}(t+\Delta t/2)
\end{eqnarray*}
with a $\Delta t$ step size and a trajectory length $\tau$. At $t=0$ $su(3)$ momenta are taken to be Gaussian $su(3)$ algebras: given eight independently distributed standard Gaussian variables at each lattice site and direction the routine \verb+algebra_su3+ computes the corresponding momentum matrices. Note the half step updates of guage fields: it is expected that the force requires more flops than the exponentiation.

The algorithm ends by correcting for the non-conservation of the Hamiltonian using Metropolis et.al. with acceptance probability:
$$
P_{\text{acc}}(\{P(0),U(0)\}\rightarrow \{P(\tau),U(\tau)\})
=\min\left\{1,e^{-[H(\tau)-H(0)]}\right\}
$$
Upon rejection, one goes back to $t=0$ and refreshes momenta. This ends the description of the Hybrid Monte Carlo algorithm. One implicit and important assumption of QCDLAB is that the \verb+rand+ function of GNU Octave suffices its purpose. The simulation routine of QCDLAB is:
\begin{small}
\begin{verbatim}
function [acc,Plaq,U1,stat]=SU3(NMC,U1,iconf);
%globals
global beta N N1 N2 N3 N4 E
beta=5.7; N1=6; N2=6; N3=6; N4=12; N=N1*N2*N3*N4;
E=ShiftOperators(N1,N2,N3,N4);
% Starting configuration
if (iconf~=2),
 U1=GaugeField(iconf); %iconf=0/1 (cold/hot)
endif
% Start Hybrid Monte Carlo
ntest=0; Plaq=[]; stat=[]; acc=0;
NMD=20; deltat = 0.025;
for mc = 1:NMC;
  p=randn(8,4*N); % Refresh momenta
  P=algebra_su3(p);
% Compute H1
  H1=T(P)-beta/3*Plaquette(U1);
% Propose U2 using MD evolution
  U2=U1;
  % MD loop
  for md=1:NMD;
    U2=MultSU3(Exp_su3(P*deltat/2),U2); % Advance fields half step
    P=P+Force_su3(U2)*deltat; % Advance momenta full step
    U2=MultSU3(Exp_su3(P*deltat/2),U2); % Advance fields half step
  end
% Compute H2
  H2=T(P)-beta/3*Plaquette(U2);
% Metropolis test
  R=min([1,exp(-(H2-H1))]);
  random=rand;
  istat=[random,R,H2-H1]; stat=[stat;istat];
  if random<R,
    U1=U2;
    acc=acc+1;
    plaq=Plaquette(U1)/N/6/3; Plaq = [Plaq;plaq];
  end
end
acc=acc/NMC
\end{verbatim}
\end{small}

Now we have enough programs to start exploring QCD. In the next section we begin with the string tension computation.

\section{QCD string}

In QCD, the energy between two static charges at large enough separation $R$ follows the string law:
$$
V(R)=KR\ ,
$$
where $K$ is the string tension. On the lattice we measure the dimensionless string tension ${\hat K}=a^2K$, where $a$ is the lattice spacing. Using the string tension value from the Regge slopes, $K=(440\text{ MeV})^2$, one can set the physical scale:
$$
a=\frac{197\text{ MeV fm}}{440\text{ MeV}}~\sqrt{\hat K}\ ,
$$
where $197\text{ MeV fm}=1$ is the energy-length conversion factor. Scale setting can be performed using any other physical quantity such as a hadron mass. QCD is compared to experiment by extrapolating dimensionless ratios of physical quantities at vanishing lattice spacing keeping the physical lattice size large enough to fit the physics. But how do we measure the string tension? We use an important lattice observable, the Wilson loop, which we deal with next.

\subsection{Wilson loop}

We already know how to measure a $1\times 1$ Wilson loop, or the plaquette. Alghough one can measure all sorts of Wilson loops on the lattice, we restrict ourseleves to rectangle Wilson loops of dimensions $R\times T$. Let $V_1$ and $V_2$ be the product of matrices along the $R$ and $T$ directions respectively:
$$
V_1=\underbrace{(U_1E_1)\cdots(U_1E_1)}_{R\text{ times}}\ ,~~~~~~~~V_2=\underbrace{(U_2E_2)\cdots(U_1E_2)}_{T\text{ times}}\ .
$$

\begin{picture}(100,180)(0,0)
\put(150, 40){\vector( 1, 0){100}}
\put(250, 40){\vector( 0, 1){100}}
\put(250,140){\vector(-1, 0){100}}
\put(150,140){\vector( 0,-1){100}}
\put(120, 30){$(0,0)$}
\put(250, 30){$(R,0)$}
\put(250,140){$(R,T)$}
\put(120,140){$(0,T)$}
\put(190, 20){$V_1$}
\put(260, 80){$V_2$}
\put(190,150){$V_1^*$}
\put(120, 80){$V_2^*$}
\end{picture}

\noindent
Then, the (unnormalized) $R\times T$ Wilson loop is:
$$
{\cal W}(R,T)=\sum_{\mu\neq\nu}\text{Re~tr~}V_1V_2V_1^*V_2^*\ .
$$
The foolowing routine is a direct implementation of the formula.
\begin{small}
\begin{verbatim}
function w=Wloop(R,T,U,N1,N2,N3,N4);
% computes rectangular Wilson loop
E=ShiftOperators(N1,N2,N3,N4);
N=N1*N2*N3*N4;
w=0;
for mu=1:4;
for nu=1:4;
  if (mu!=nu),
    E1=E(:,(mu-1)*3*N+1:mu*3*N); E2=E(:,(nu-1)*3*N+1:nu*3*N);
    U1=U(:,(mu-1)*3*N+1:mu*3*N); U2=U(:,(nu-1)*3*N+1:nu*3*N);
    U1=U1*E1; U2=U2*E2;
    V1=speye(size(U1));
    for r=1:R;
      V1=V1*U1;
    end
    V2=speye(size(U2));
    for t=1:T;
      V2=V2*U2;
    end
    w=w+real(trace(V1*V2*V1'*V2'));
  end
end
end
\end{verbatim}
\end{small}

\subsection{Area law}

The vacuum expectation value of $R\times T$ Wilson loop is the correlation function of the static quark-antiquark propagator separated with $T$ lattice sites. Since, the large time behavior is dominated by the ground state contribution:
$$
W(R,T)=\langle{\cal W}(R,T)\rangle\simeq e^{-V(R)T}\ ,
$$
the string behavior of the potential $V(R)=KR$ is observed if the Wilson loop falls off exponentially with the area of the loop $RT$. A direct way to measure the string tension is by means of Creutz ratios:
$$
\chi(R,T)=-\log\frac{W(R+1,T+1)W(R,T)}{W(R+1,T)W(R,T+1)}
$$
at large $R$ and $T$ values. Note that $W$'s are sample averages and error propagation is not straightforward. A proper way to compute the error is using partial sample averages. This requires a large sample volume, which is often not available. An important trick, used as a short cut, is data resampling, or the so called {\it bootstrap} resampling. We will expain shortly, a bootstrap variant which is widely used in lattice QCD, the {\it jackknife} method.

\subsection{Jackknife resampling}

Let us suppose we are given the real data vector $x$ with volume $n$. The resampled jackknife data is the linear map:
$$
x^{(J)}=\bar{x}-\frac{x-\bar{x}}{n-1}\ ,
$$
which conserves the sample average $\bar x$. If we require further the conservation of variance:
$$
{\text{Var}}\left[x^{(J)}\right]={\text{Var}}(x)~~~~~~~~\Leftrightarrow~~~~~~~~C\norm{x^{(J)}-\bar{x}}^2=\frac{1}{n-1}\norm{x-\bar{x}}^2\ ,
$$
we should should choose the normalization factor $C=n-1$. As it is clear by inspection, the elements or $x^{(J)}$ are partial sample averages of $x$:
$$
x_i^{(J)}=\frac{1}{n-1}\sum_{k(\neq i)}x_k\ .
$$
This way, we gain $n-1$ more sample averages of primary data than in the case without resampling. Thus, the physical quantity of interest, for example a Creutz ratio, is computed for {\it all} individual elements of $x^{(J)}$ as opposed to a single sample average that was available originally. If $y$ is the vector such estimations, its sample variance normalization is inherited from the corresponding variance of resampled data:
$$
{\text{Var}}(y)=(n-1)\norm{y-\bar{y}}^2\ .
$$
The QCDLAB routine that computes Creutz ratios is \verb+creutz_ratios+. Applying it to a small sample of 40 Wilson loops of maximal linear size 4, obtained on $6^3\times 12$ lattices, we get:
$$
\chi(2,2)=0.372(3)\ ,~~~~~~~~\chi(3,3)=0.26(2)\ ,~~~~~~~~\chi(4,4)=0.2(1)
$$
with the corresponding estimation of the lattice spacings, in fm units:
$$
a_{(2,2)}=0.273(1)\ ,~~~~~~~~a_{(3,3)}=0.23(1)\ ,~~~~~~~~a_{(4,4)}=0.19(7)\ .
$$
Normally we should rely on the results of large Wilson loops. We see however, that the $a_{(4,4)}$ value has a large error, so that the compromise is to select the $a_{(3,3)}$ value as the estimation of our scale.

\subsection{Quark-antiquark potential}

A standart way to measure the string tension on the lattice is measuring the quark-antiquark potential. In order to extract the potential from the Wilson loops one usually relies on {\it effective potentials}:
$$
V_{\text{eff}}(R,T)=-\log\frac{W(R,T+1)}{W(R,T)}\ .
$$
For fixed $R$ we select as $V(R)$ the median value of $V_{\text{eff}}(R,T)$ over all $T$ sizes of Wilson loops, whereas the corresponding error is computed using the jackknife method. These data are fitted to the general form:
$$
V(R)=V_o+\frac{\alpha}{R}+KR\ ,
$$
where $V_o$ and $\alpha$ are two more constants in addition to the string tension $K$. Such a procedure is coded in the routine \verb+effective_potentials+. Using the same set of Wilson loops as in the case of Creutz ratios the routine produces the 3-sigma band plot:

\begin{center}
\includegraphics[scale=0.50]{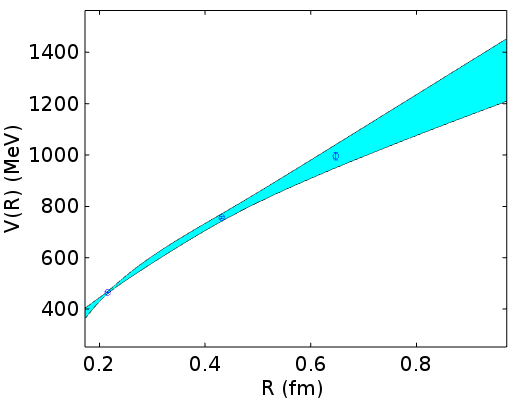}
\end{center}

\noindent
as well as the results:
$$
\chi=0.23(2)\ ,~~~~~~~~~~~~a_{\text{pot}}=0.22(1)\text{~fm}\ .
$$
From the plot we read that the potential is around 1.2 GeV if it is extrapolated at 1 fm separation. We observe also that the results overlap within 1-sigma with those obtained using Creutz ratios.

\section{Hadron spectrum}

A basic computation in lattice QCD is the hadron spectrum. We will ilustrate it in the case of low lying mesons such as the pion and rho. Quark propagators are computed using the \verb+quark_propagator+ routine. It calls the BiCGstab algorithm \cite{van_der_vorst,gutknecht} as a Dirac solver:
$$
q=D^{-1}b\ ,
$$
where $D$ is the Wilson operator and $b$ a point source at the origin of the lattice for each color and Dirac spin. Therefore, at each lattice site $x$ the propagator $q_x$ is a $12\times 12$ matrix. The pion and rho propagators are defined as:
$$
G_{\pi,x}=\text{Tr~}q_xq_x^*\ ,~~~~~~~~G_{\rho,x}=\text{Tr~}\gamma_5\gamma_kq_x\gamma_k\gamma_5q_x^*\ ,
$$
where the trace and Hermitian conjugation is performed in the tensor product space of color and spin. We sum over space-like lattice sites in order to get particle masses. For example, at long Euclidean time separation $T$ the pion propagator is dominated by its ground state contribution:
$$
G_{\pi,T}=\sum_{\vec{x}}G_{\pi,(T,\vec{x})}\simeq Ce^{-m_\pi T}\ .
$$
Since we simulate with periodic boundary conditions in all directions the propagator decays exponentially also with respect to reflected times, which are translated by the lattice size $N_4$:
$$
G_{\pi,T}\simeq C\left[e^{-m_\pi T}+e^{-m_\pi(N_4-T)}\right]\propto\cosh\left(T-\frac{N_4}{2}\right)\ .
$$
Therefore, the routine \verb+pion_propagator+ symmetrizes propagators with respect to the origin, which is actually at $T=1$:
\begin{small}
\begin{verbatim}
N1=6;N2=6;N3=6;N4=12;
N=N1*N2*N3*N4;
pion=sum(abs(q).^2,2);
pion=sum(reshape(pion,12,N));
pion=sum(reshape(pion,N4,N1*N2*N3),2);
pion(2:N4/2)=(pion(2:N4/2)+pion(N4:-1:N4/2+2))/2;
pion(N4/2+2:end)=[];
\end{verbatim}
\end{small}

\subsection{Effective masses}

In complete analogy to the quark-antiquark potential we compute the effective masses of mesons as:
$$
M_{\text{eff}}(T)=-\log\frac{G_{\pi,T+1}}{G_{\pi,T}}
$$
and take the median value over all $T$ values as the actual $M_{\text{eff}}$. The meson masses squared are then fitted against the bare quark masses using a linear model:
$$
M_{\text{eff}}^2=c_o+c_1m\ ,
$$
where $c_o,c_1$ are unknown constants. With Wilson fermions we define the chiral limit at the vanishing pion mass. This procedure can be implemented by first calling the routine \verb+effective_pion_masses+ with pion data to find the critical quark mass $m_c$. Then, the routine \verb+effective_rho_masses+ is called with rho data and critical quark mass as input. It returns the lattice spacing using $M_\rho=770\text{~MeV}$. Finally, the routine \verb+effective_pion_masses+ is called once more with pion data and the lattice spacing as input. Using the same configurations as before we get the 1-sigma band plot for pion and rho masses:

\begin{center}
\includegraphics[scale=0.50]{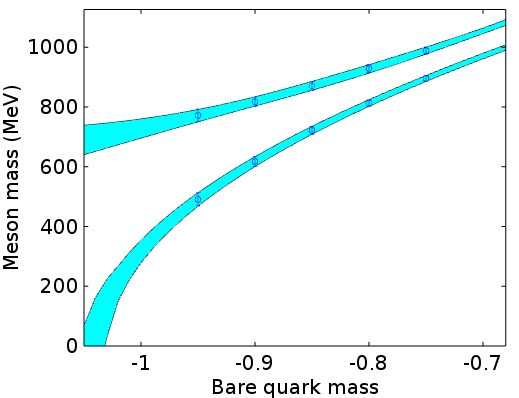}
\end{center}

\noindent
We have used a quadratic fit for the rho mass. The plot shows that the pion mass squared vanishes linearly with the quark mass in the chiral limit:
$$
M_\pi^2=c_1(m-m_c)\ .
$$
The nonzero value of $m_c$ is an artifact of Wilson fermions. For chirally symmetric fermions $m_c$ should be zero. We these data, the estimated lattice spacing:
$$
a_\rho=0.23(1)\text{~fm}
$$
and the one estimated using the quark-antiquark potential overlap within 1-sigma errors.

\section{Autocorrelations}

An important issue that one must take into the consideration in error reporting are autocorrelations in the Monte Carlo time series. If $x^{(1)}=({\cal O}_1,{\cal O}_2,\ldots,{\cal O}_n)$ is the time series vector of an observable $\cal O$ one measures the autocorrelation function between $x^{(1)}$ and the time forwarded samples $x^{(1)},x^{(2)},\ldots,x^{(t)}$:
$$
f_{j,{\cal O}}=C\sum_{k=1}^{n-t+1}\left(x_k^{(1)}-\bar{x}^{(1)}\right)\left(x_k^{(j)}-\bar{x}^{(j)}\right)\ ,~~~~~~~~j=1,2,\ldots,t
$$
as well as the {\it integrated autocorrelation time} $\tau_{\text{int},{\cal O}}$ \cite{sokal}:
$$
\tau_{\text{int},{\cal O}}=\frac{1}{2}+\sum_{j=2}^n\left(1-\frac{j-1}{n}\right)f_{j,{\cal O}}\ .
$$
The normalization constant $C$ is chosen such that $f_1=1$. The right hand side may be approximated by the sum:
$$
\tau_{\text{int},{\cal O}}\approx\frac{1}{2}+\sum_{j=2}^{t}f_{j,{\cal O}}
$$
assuming that the data volume is much larger than the cutoff $t$. QCDLAB routine that computes autocorrelations is \verb+Autocorel+:
\begin{small}
\begin{verbatim}
function [tau_int,f]=Autocorel(x,t);
% x: data vector of length N
% t: forward time
% tau_int: integrated autocorrelation time
% f: autocorrelation function
x=x(:);
N=max(size(x));
x1=x(1:N-t+1);
x1=x1-mean(x1)*ones(N-t+1,1);
f=zeros(t,1);
for j=1:t;
  xj=x(j:N-t+j);
  xj=xj-mean(xj)*ones(N-t+1,1);
  f(j)=x1'*xj/(N-t+1);
end
f=f/f(1);
tau_int=1/2+sum(f(2:t));
\end{verbatim}
\end{small}
A proper estimation of autocorrelations should also ensure that the integrated autocorrelation times are small compared to $t$, and the corresponding error is computed on a large data set. In our simulation example, plaquette decorrelates in $5(2)$ Hybrid Monte Carlo trajectories and saved configurations are separated by 100 trajectories.

\section{Minimally doubled chiral fermions}

In this section we describe new features included in the version 2.1 of QCDLAB. When studying the spontaneous breaking of chiral symmetry with Wilson fermions we encounter a nonzero critical bare quark mass, which is a consequence of the explicit breaking of the chiral symmetry of the Wilson-Dirac operator. One way to circumvent this problem si by using Ginsparg-Wilson fermions which posses exact chiral symmetry on the lattice \cite{neuberger}. Such fermions, which are part of the QCDLAB versions 1.0 and 1.1  \cite{qcdlab1}, are compuationally complex and we skip them in the follwing. Instead, we use here chiral fermions with broken hypercubic symmetry \cite{kw,bc}. These fermions describe a degenerate isospin doublet and are ideal for studying QCD with $u$ and $d$ quarks. However, due to broken hypercubic symmetry they lead to counterterms in the action \cite{capitani_et_al}. We use the version of reference \cite{bc} since it has a negligibly small gauge action counterterm, which we set to zero as a first approximation. The structure of the Dirac operator resembles the one of the Wilson operator:
$$
D_{BC}(U)=mI+i(-2+c_3)\Gamma+\frac{1}{2}\sum_{\mu}[(U_\mu E_\mu)\otimes(\gamma_\mu+i\gamma_\mu')+(U_\mu E_\mu)^*\otimes(-\gamma_\mu+i\gamma_\mu')]\ ,
$$
where the new set of gamma matrices $\gamma_\mu'=\Gamma\gamma_\mu\Gamma$ is associated to the doublet partner fermion and $\Gamma=\sum_\mu\gamma_\mu/2=\sum_\mu\gamma_\mu'/2$. Apart from a negligibly small counterterm, which we have set to zero, the counterterm $ic_3\Gamma$ is used to restore the hypercubic symmetry of the theory. In the follwoing figure we see that the neutral pion mass, the only Goldstone boson of the theory, flattens at $c_3=0.2$ (we have fixed the bare quark mass at the small value $0.01$):

\begin{center}
\includegraphics[scale=0.4]{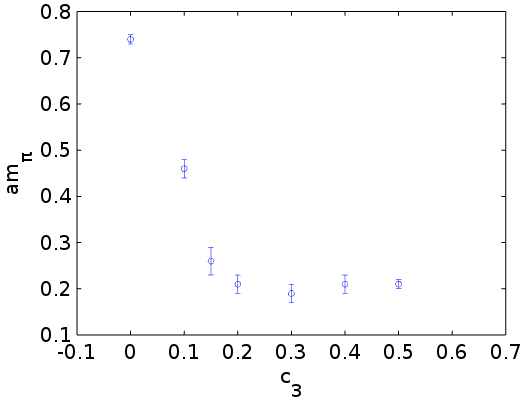}
\end{center}

\noindent
We have used the same meson operators as with Wilson fermions keeping in mind the two-fold degeneracy of the isospin doublet fermion. For example, if the fermion doublet $\psi$ is written in terms of flavor singlet fields $u$ and $\Gamma d$, i.e. $\psi=u+\Gamma d$, the pion operator admits the expression of the neutral pion operator:
$$
\bar\psi\gamma_5\psi=\bar u\gamma_5 u+\bar d~\Gamma\gamma_5\Gamma d=\bar u\gamma_5 u-\bar d\gamma_5d
$$
($\Gamma^2=1$ and it anticommutes with $\gamma_5$). Note that cross terms vanish identically since $u$ and $d$ are defined on the orthogonal subspaces of the flavor doublet field $\psi$. With the value of $c_3$ being fixed at $0.2$ one may compute the pion and rho masses for a range of bare quark masses. Below we plot the fitted data with the 1-sigma band:

\begin{center}
\includegraphics[scale=0.5]{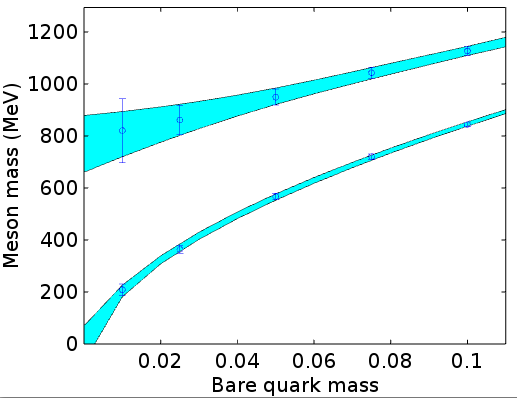}
\end{center}

\noindent
Note that we have included a quenched chiral logarithm in the pion mass model:
$$
M_\pi^2=c_1~m+c_2~m\ln m\ ,
$$
since it gives a better fit of the data. The lattice spacing value $a_\rho=0.20(3)~\text{fm}$ overlaps with the value found using Wilson fermions within the 1-sigma error.

\bigskip
In summary, we have described briefly the main routines of QCDLAB, versions 2.0 and 2.1, as well as its use.

\subsection*{Acknowledgement}

The author thanks Travis Whyte who found a missmatch of shift operators at the \verb+su3+ directory and inside the \verb+Wilson.m+ function.

\end{document}